\documentstyle[tighten,aps,axodraw]{revtex}

\twocolumn

\newcommand{\be}{\begin{equation}}
\newcommand{\ee}{\end{equation}}
\newcommand{\ba}{\begin{eqnarray}}
\newcommand{\ea}{\end{eqnarray}}

\def\vec#1{{\mbox{\boldmath$#1$}}}

\newcommand{\r}{\mbox{$\vec{r}$}}

\newcommand{\ep}{\epsilon}

\begin{document}

\title{
%
%
\[ \vspace{-2cm} \]
\noindent\hfill\hbox{\rm  SLAC-PUB-8281, TTP99-44} \vskip 1pt
\noindent\hfill\hbox{\rm hep-ph/9911277} \vskip 10pt
%
%
The three loop slope of the Dirac form factor and the $1S$ Lamb shift
in hydrogen}

\author{Kirill Melnikov\thanks{
e-mail:  melnikov@slac.stanford.edu}}
\address{Stanford Linear Accelerator Center\\
Stanford University, Stanford, CA 94309}
\author{Timo van Ritbergen\thanks{
e-mail:  timo@particle.physik.uni-karlsruhe.de}}
\address{Institut f\"{u}r Theoretische Teilchenphysik,\\
Universit\"{a}t Karlsruhe,
D--76128 Karlsruhe, Germany}
\maketitle

\begin{abstract}
The last unknown contribution 
to hydrogen energy levels at order $m\alpha ^7$,
due to the slope of the Dirac form factor
at three loops, is evaluated in a closed analytical form. The resulting shift
of the hydrogen $n$S  energy level is found
to be $3.016/n^3~{\rm kHz}$. 
Using the QED calculations of the $1S$ Lamb shift,
 we extract a precise value of the proton charge 
 radius $r_p=0.883 \pm 0.014~{\rm fm}$.

\end{abstract}


Precision experiments with hydrogen and,
more generally, with hydrogen-like atoms 
serve as an excellent laboratory to test theoretical  
approaches to bound state QED
(for a recent review see, e.g. \cite{kinoshita}). 
These experiments address a number of features 
of the simplest atoms, such as the energy levels 
of the ground and excited states, and the corresponding lifetimes.

In recent years we have seen remarkable progress
in the experimental study of the hydrogen atom. 
In particular the accuracy of the $1$S
Lamb shift measurements has increased dramatically
over the years \cite{weitz,berkelend,bourzeix,beauvoir,udem,schwob}. 
All measurements are consistent with each other and the most accurate value
so far was determined in Ref.\cite{schwob}:
\be
\Delta E(1S)_{\rm exp} = 8~172~837(22)~{\rm kHz}.
\ee
This result was obtained by analysing the most precise
measurements for the transition frequencies in hydrogen
(see \cite{schwob} for details).

Theoretically, 
since the ratio of the electron mass $m$  to the proton mass $M$ 
is very small, it is convenient to write hydrogen energy 
levels as a double expansion  in $\alpha$ and $m/M$.
The corrections which survive in the limit $M \to \infty $
are known as non-recoil corrections, the other corrections are called
recoil ones. It is further convenient to organize non-recoil
corrections in powers of $\alpha^n (Z\alpha)^l$, assigning 
an auxiliary notation $Z$ for the proton charge. In this 
case the correction $\alpha^n (Z\alpha)^l$ describes a contribution 
of all diagrams with $l-3$ Coulomb photons exchanged between 
the electron and the proton and with $n$ photons emitted 
and absorbed by the electron. The general expression for 
the $n$S-level shift can be written as:
\be  \label{deltaEsplit}
\Delta E = \Delta E_{\rm non-recoil} + \Delta E_{\rm recoil}
 +\Delta E_{\rm vp} + \Delta E_{\rm proton },
\ee
where we have also added a contribution of muons and 
hadrons to photon vacuum polarization 
and  a contribution due to the proton structure
(see a discussion below).

We begin  with non-recoil corrections. These corrections 
can be parameterized as (see e.g. \cite{sapirstein,mohr,preview}):
\ba
\Delta E_{\rm non-recoil} &&= \frac {m \alpha (Z\alpha)^4}{\pi n^3}
 \left (\frac {m_r}{m} \right )^3 
 \Large ( A_{40} + A_{41} L
 +(Z\alpha) A_{50} 
 \nonumber \\ &&
+ (Z\alpha)^2 \left [
 A_{62} L^2
+ A_{61} L
+ A(Z\alpha)
\right ] \Large )
\nonumber \\
&&
 +\frac {m \alpha^2 (Z\alpha)^4}{\pi^2 n^3} 
 \left (\frac {m_r}{m} \right )^3
  \left ( B_{40} 
 + (Z\alpha) B_{50} \right )
\nonumber \\
&&
 + \frac {m \alpha^3 (Z\alpha)^4} {\pi^3 n^3} 
  \left (\frac {m_r}{m} \right )^3 C_{40}
 +{\cal O}(m \alpha^8 \log^3 \alpha). 
\label{energy}
\ea
where $m_r = mM/(m+M)$ is the reduced mass of the electron and
$L = \log m/(m_r (Z\alpha)^2)$ and the function $A(Z\alpha)$
contains all higher order terms in the expansion in $Z\alpha$.
We have also indicated that 
the higher order corrections contain a logarithm of the 
fine structure constant in the third power \cite{Karsh}. 

All the terms in the above equation, with the exception for $C_{40}$,
are currently known. It is the purpose of this paper to report 
on the calculation of the last missing ingredient in $C_{40}$,
the slope of the Dirac form factor at zero momentum transfer.

The hydrogen atom is formed because of a Coulomb interaction 
of the proton with the electron. The interaction of the 
virtual photon with the electron on its mass shell  
can be parameterized by the so-called Dirac and Pauli  form factors: 
$$
\bar u(p_2) \Gamma_\mu u(p_1) =
\bar u(p_2) \left ( F_1(q^2) \gamma _\mu 
 + i \sigma_{\mu \nu} \frac {q_\mu}{m} F_2(q^2) \right ) u (p_1),
$$
where the $u(p)$ are the electron spinors in the initial and 
final state and $q$ is the momenta carried away by the photon.
The momenta satisfy the relation $q = p_2 - p_1$.
An important consequence of QED gauge invariance and the 
electron charge definition is that the Dirac form factor 
equals unity at zero momentum transfer
$F_1(0) = 1$, to all orders in the coupling constant. 
The Pauli form factor at zero momentum transfer describes 
an  interaction of the electron spin  with the homogeneous 
magnetic field; it is the electron anomalous magnetic moment.

Let us now turn to the contribution of the Dirac form factor 
to hydrogen energy levels. The typical momenta of the 
Coulomb photon in hydrogen is given by the inverse Bohr radius
$q_{\rm typ} \sim m \alpha$, where $m$ is the electron mass.
Compared to the electron mass, this momentum transfer is quite 
small  and for this reason one can Taylor expand 
the Dirac form factor in powers of $q^2/m^2$. One obtains:
$$
F_1(q^2) = 1 + F_1'\frac{q^2}{m^2} 
    + {\cal O}\left( \frac{q^4}{m^4} \right) ,
$$
where, as we already mentioned, the first term equals 
unity to all orders in the coupling constant. The slope of the Dirac 
form factor, $F_1'$ in the above formula, can be written 
as a series in $\alpha$:
\be
 F_1' = \sum \limits_{n=1}^{\infty}
 \left (\frac {\alpha}{\pi} \right )^{n}
 A^{(n)}_{\rm slope}.    
\label{slope}
\ee
In this paper we consider the first three terms in this series.

Two things should be noted at this point. The 
first and second order corrections to the slope 
of the Dirac form factor contribute to the coefficients
$A_{40}$ and $B_{40}$ in Eq.(\ref{energy}), respectively.
Moreover, the one loop slope $A_{\rm slope}^{(1)}$ is not infrared 
finite, as can be seen from the fact that the 
term $A_{41} \log(\alpha)$ appears in Eq.(\ref{energy}).
This infrared divergence gets removed if one takes
into account that the electron is not on its mass shell
in the bound state. In the language of the 
effective theories, the one loop slope in Eq.(\ref{slope})
corresponds to ``hard'' contributions to energy levels.
Since the two and three loop slopes of the Dirac form factor
are infrared finite, the off-shellness of the electron is irrelevant
for these contributions.
 
Being divergent, the one loop slope depends upon the chosen 
regularization.  In the present calculation 
we have used dimensional regularization
(the space-time dimension is $D=4-2\ep$) for both ultraviolet
and infrared divergences. 

Let us briefly describe how the actual calculation of $A_{\rm slope}^{(3)}$
has been 
done. After applying a projection operator for
 the Dirac form factor on the electron-photon vertex
 and after performing the Taylor expansion in the photon 
momentum transfer, one obtains diagrams of the self energy 
type. There are four  basic topologies which appear in this calculation, 
characteristic example diagrams are depicted in Fig.1. 
 For each of the topologies 
one writes down a system of recurrence relations obtained
by the use of integration-by-parts \cite{ibp}.  Solving this system, it is 
possible to show that any integral which belongs to the  
above topologies can be expressed through $17$ master integrals.
Luckily, these seventeen integrals have already been 
computed in the course of the analytical calculation 
of the electron anomalous magnetic moment 
\cite{Laporta} and hence can be taken from there.
Let us mention, that as a check of the calculation, and in particular 
 on our solution of the system of 
recurrence relations, we have reproduced also the analytical value of 
the three-loop electron anomalous magnetic moment.
Details of our calculation, including the results for the individual 
diagrams, will be  presented in a separate publication.

Our result for the slope of the Dirac form factor is
\footnote{The first  correct numerical result on the two loop 
slope of the Dirac form factor was published in Ref.\cite{brodsky}.
The analytical result was obtained in \cite{barbieri}.}:
\ba
A_{\rm slope}^{(1)} = &&   - \frac {1}{8} - \frac {1}{6 \ep} ,
\\
A_{\rm slope}^{(2)} =&&   - \frac {4819}{5184} - \frac {3}{4}\zeta_3 
 + \frac {1}{2}\pi^2 \log 2 - \frac {49}{432}\pi^2,
\\
A_{\rm slope}^{(3)}
          = && - \frac {77513}{186624} - \frac {17}{24} \zeta_3 \pi ^2 
         - \frac {2929}{288} \zeta_3 + \frac {25}{8}\zeta _5 
\nonumber \\
         && + \frac {41671}{2160}\pi^2 \log 2
          - \frac {103}{1080}\pi^2 \log^2 2 
         - \frac {454979}{38880}\pi ^2 
\nonumber \\
&&         + \frac {3899}{25920}\pi^4 - \frac {217}{216} \log^4 2 
         - \frac {217}{9} a_4, 
\ea
 where
  $a_4 =\sum \limits_{n=1}^{\infty} 1/(2^n n^4) $
 and $\zeta_k = \sum \limits_{n=1}^{\infty} 1/(n^k)$ 
 denotes the Riemann zeta function.

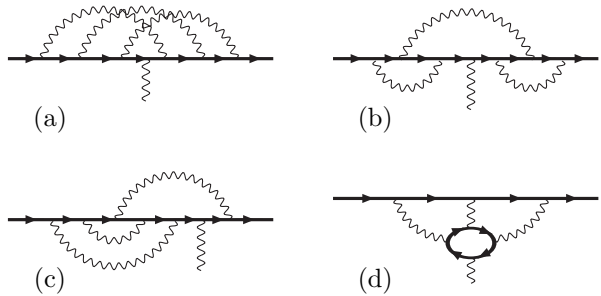
\begin{figure}
\begin{center}
\hfill
\begin{picture}(120,40)(0,0)
 \SetScale{.8}
 \SetWidth{1.5}
 \Line(5,20)(130,20)
 \SetWidth{0.8}
 \ArrowLine(10,20)(22,20)
 \ArrowLine(22,20)(42,20)
 \ArrowLine(42,20)(62,20)
 \ArrowLine(58,20)(78,20)
 \ArrowLine(80,20)(96,20)
 \ArrowLine(100,20)(110,20)
 \ArrowLine(118,20)(125,20)
 \SetWidth{0.5}
 \PhotonArc(86,14)(27,13,167){1.6}{16.5}
 \PhotonArc(68,14)(29.2,13,167){1.6}{16.5}
 \PhotonArc(50,14)(29.2,13,167){1.6}{16.5}
  \Photon(70,20)(70,0){1.8}{4}
 \Text(20,-2)[t]{(a)}
\end{picture}
\hfill
\begin{picture}(120,40)(0,0)
 \SetScale{.8}
 \SetWidth{1.5}
 \Line(5,20)(130,20)
 \SetWidth{0.8}
 \ArrowLine(10,20)(22,20)
 \ArrowLine(24,20)(40,20)
 \ArrowLine(40,20)(55,20)
 \ArrowLine(60,20)(70,20)
 \ArrowLine(70,20)(85,20)
 \ArrowLine(86,20)(96,20)
 \ArrowLine(102,20)(110,20)
 \ArrowLine(118,20)(125,20)
 \SetWidth{0.5}
 \PhotonArc(98,21)(15,188,352){1.6}{9.5}
  \PhotonArc(40,21)(15,188,352){1.6}{9.5}

 \PhotonArc(68,10)(32,17,163){1.6}{17.5}
  \Photon(70,20)(70,-4){1.8}{5}
 \Text(20,-2)[t]{(b)}
\end{picture}
\hfill\null\\
\hfill
\begin{picture}(120,60)(0,0)
  \SetScale{.8}
 \SetWidth{1.5}
 \Line(5,20)(130,20)
 \SetWidth{0.8}
  \ArrowLine(10,20)(22,20)
 \ArrowLine(25,20)(42,20)
 \ArrowLine(42,20)(62,20)
 \ArrowLine(67,20)(87,20)
 \ArrowLine(83,20)(99,20)
 \ArrowLine(98,20)(110,20)
 \ArrowLine(112,20)(125,20)
 \SetWidth{0.5}
 \PhotonArc(55,24)(14,197,343){1.6}{7.5}
 \PhotonArc(55,28)(30,197,343){1.6}{16.5}
 \PhotonArc(83,14)(27,13,167){1.6}{16.5}
  \Photon(96,20)(96,-4){1.8}{5}
 \Text(20,-2)[t]{(c)}
\end{picture}
\hfill
\begin{picture}(120,60)(0,0)
    \SetScale{.8}
 \SetWidth{1.5}
 \Line(5,30)(130,30)
  \SetWidth{1.3}
  \Oval(70,9)(7,11)(0)
 \SetWidth{0.8}
 \ArrowLine(15,30)(28,30)
 \ArrowLine(42,30)(62,30)
 \ArrowLine(83,30)(99,30)
 \ArrowLine(115,30)(120,30)

  \ArrowLine(61.5,14)(66.5,16)
  \ArrowLine(74,15.5)(80,13)
   \ArrowLine(79,5.0)(73,2)
   \ArrowLine(65,2.7)(60,5.5)

 \SetWidth{0.5}
  \Photon(70,30)(70,16){1.8}{3}
 \PhotonArc(59,36)(25,195,270){1.6}{7.5}
 \PhotonArc(81,36)(25,270,345){1.6}{7.5}
  \Photon(70,2)(70,-10){1.8}{2.5}
 \Text(20,-2)[t]{(d)}
\SetScale{1}
\end{picture}
\hfill\null\\
\vglue 18pt
\end{center}
\caption{Examples of electron-photon vertex diagrams that correspond to 
   the four different integration topologies. }
\label{fig:examplediagrams}
\end{figure}

We now turn to the contribution of the slope of the Dirac 
form factor to hydrogen energy levels. It is easy to see 
that the slope of the Dirac form factor gives rise to a
perturbation of the Coulomb potential:
$$
\delta V_{\rm slope}(r) = \frac {4\pi Z \alpha}{m^2} \delta^{3}(\r)
F_1'.
$$
This perturbation delivers a
$n$S-level energy shift:
$$
\Delta E_{\rm slope} = \langle \psi (r) | \delta V_{\rm slope}(r)
 |\psi (r) \rangle
= \frac {4\pi \alpha Z}{m^2}F_1' |\psi(0)|^2.
$$
Here $|\psi(0)|^2 = (m_r \alpha Z)^3/(\pi n^3)$ is the square of 
the hydrogen wave function at the origin. The correction
induced by the three-loop slope of the Dirac form factor 
is then\footnote{In what follows numerical results are given in frequency units
using $\Delta E \to \Delta E/(2\pi \hbar)$.}:
\ba
\Delta E_{\alpha^3(Z\alpha)^4}^{\rm slope}(n) 
&& = \frac {4 m \alpha^7}{\pi^3 n^3}
 \left ( \frac {m_r}{m} \right )^3 
 A_{\rm slope}^{(3)} 
\nonumber \\
 && \to  
 \frac {8 c R_{\infty} \alpha^5}{\pi^3 n^3} \left ( \frac {m_r}{m} \right )^3
 A_{\rm slope}^{(3)},
\ea
where $c R_{\infty}$ is the Rydberg constant in MHz.

Using the  values for the Rydberg  and the 
fine structure constant \cite{nist} 
\ba
&&c R_{\infty} = 3~289~841~960.367(25)~{\rm MHz},
\nonumber \\
&& \alpha = 1/137.035~999~76(50),
\ea
we arrive at:
\be
\Delta E_{\alpha^3(Z\alpha)^4}^{\rm slope}(n) = \frac {3.016}{n^3}~{\rm kHz}. 
\ee

The contribution due to the slope of the Dirac form factor was the 
last unknown contribution to the hydrogen energy levels at 
order $\alpha^3 (Z\alpha)^4$. The two other contributions to
the coefficient $C_{40}$
come from the three-loop electron anomalous magnetic moment
and the three-loop vacuum polarization correction to the Coulomb
propagator. These contributions can be extracted from the 
literature \cite{Laporta,Baikov}.

Taking all three contributions into account, 
  we obtain the following expression for the coefficient
$C_{40}$ for the S-levels:
\ba
C_{40} &&= 
\frac {679441}{93312}-\frac {252251}{9720}\pi^2
+\frac {4787}{108}\pi^2 \log 2
\nonumber \\
&& -\frac {121}{72} \zeta _3 \pi^2
 -\frac {239}{135} \pi^2 \log^2 2
 -\frac {84071}{2304}\zeta _3
+ \frac {85}{24}\zeta_5 
\nonumber \\
&&
 - \frac {568}{9}a_4
 +\frac {1591}{3240}\pi^4 - \frac {71}{27}\log ^4 2 \approx
0.417508,
\ea
which results in a three-loop correction to the $n$S-level Lamb shift:
\ba
&&\Delta E_{ \alpha^3 (Z\alpha)^4 }(n)  = 
\frac {m \alpha^7}{n^3 \pi^3}
\left ( \frac {m_r}{m} \right )^3 C_{40}
\nonumber \\
&&= \left( 3.016 + 5.187 - 6.370 \right )~\frac{{\rm kHz}}{n^3}
= \frac {1.83}{n^3}~{\rm kHz}.
\ea

In the above equation we have displayed the  
 contributions due to the three-loop
slope of the Dirac form factor, the three-loop anomalous magnetic 
moment of the electron and the three-loop photon vacuum polarization
function separately to emphasize a strong cancellation which 
occurs between vertex and vacuum polarization corrections.
Thanks to this cancellation, the correction turns out to be quite 
small numerically.

We now turn to  other results available in the literature.
For the coefficients $A_{ij}$ and $B_{ij}$ in Eq.(\ref{energy})
we use\footnote{See Ref.\cite{sapirstein} for reference to earlier work.
Some of the numbers in  Ref.\cite{sapirstein} have been updated 
according to recent calculations in 
Refs.\cite{Pachucki1,Pachucki2,Eides1}. The value of the non-perturbative
function $A(1S)(Z\alpha)$ for $Z=1$ is extracted from 
Ref.\cite{jent} for the self energy correction and from 
Ref.\cite{mohr} for the vacuum polarization.}
\ba
&&A_{40}(nS) = \frac {10}{9} -\frac {4}{15} - \frac {4}{3} \log k_0(nS),~~~
A_{41}(nS) = \frac {4}{3},
\nonumber \\
&&\log k_0(1S) = 2.9841285557655(1),
\nonumber \\
&& A_{50}(nS) = \pi \left ( \frac {139}{32} + \frac {5}{48} - 
2 \log 2 \right ),~~~
A_{62}(nS) = -1,
\nonumber \\
&&
A_{61}(1S) = \frac {28}{3} \log 2 - \frac {21}{20} -\frac {2}{15},
\nonumber \\
&&A(1S)(Z\alpha) = -30.29024(2) 
- -0.6187
+ \left ( \frac {19}{45}-\frac {\pi^2}{27} \right ),
\nonumber \\
&&B_{40}(nS) =  -\frac {2179}{648}-\frac {9}{4}\zeta_3
 +\frac {3}{2}\pi^2 \log 2 - \frac {10}{27}\pi^2,
\nonumber \\
&&B_{50}(1S) =  -21.558(3).
\nonumber 
\ea

The only known non-recoil correction at order ${\cal O}(m\alpha^8)$ 
is the triple logarithmic enhanced contribution \cite{Karsh}:
$$
\Delta E_{\alpha^2(Z\alpha)^6} 
 = -\frac {8}{27} \frac {m\alpha^2 (Z\alpha)^6}{\pi^2 n^3} 
\log^3 \frac {1}{(Z\alpha)^2} = -\frac {28.4}{n^3}~{\rm kHz}.
$$
This correction is included in the central value for the $1$S Lamb
shift quoted below.

Consider now the recoil corrections. Part of these are already included
in Eq.(\ref{energy}) through its dependence on the reduced mass
of the electron. The terms which go beyond this approximation 
are:
\ba
\Delta E_{\rm recoil} && = \frac {m^2(Z\alpha)^5}{\pi M}
\left ( \frac {1}{3} L -\frac {8}{3}\log k_0(1S)
+\frac {62}{9} + \frac {14}{3} \log 2
\right. \nonumber \\
&&\left. - \frac {2}{M^2 - m^2} \left ( M^2 \log \frac {m}{m_r} -
 m^2 \log \frac {M}{m_r} \right )  \right )
\nonumber \\
+ && \frac  {m^2(Z\alpha)^6}{ M} \left ( 4\log 2 - \frac {7}{2} \right )
- \frac {\alpha  (Z\alpha)^5 m^2}{\pi^2 M}
 \left ( 1.36449(2) \right ).
\nonumber 
\ea
computed in Refs.\cite{Salpeter,Pachucki3,Pachucki4}.

The vacuum polarization effects due to muons and hadrons can 
be extracted from Ref.\cite{friar} and give
$\Delta E_{\rm vp} = -8.5~{\rm kHz}$ for the $1$S Lamb shift.

The last contribution in Eq.(\ref{deltaEsplit}) which has been 
 not considered so far is the one due to proton structure.
The major part of this correction to the Lamb shift 
is parameterized through the proton charge radius but there 
is also a proton self energy correction which goes beyond
that approximation \cite{Pachucki4}.
The complete correction reads:
\ba
\Delta E_{\rm proton} = &&\frac {2(Z\alpha)^4}{3n^3} 
\langle r_p^2 \rangle  m_r^3
+\frac {4m_r^3}{3\pi n^3 M^2} (Z^2\alpha)(Z\alpha)^4
\nonumber \\
&& \times \left [ \log \frac {M}{m_r(Z\alpha)^2} 
- \log k_0(n) \right ].
\ea
There are two different values for the proton charge radius
$r_p =0.805(11)~{\rm fm}$ and $r_p=0.862(12)~{\rm fm}$ obtained 
in \cite{hand,simon}, respectively, by analysing the electron proton
scattering data.  The smaller value, $r_p =0.805(11)~{\rm fm}$,
seems to give a serious disagreement between precision atomic
measurements and the theoretical predictions \cite{karsh2}. 
For this reason we do not consider it here. 
The data of Ref. \cite{simon}, on the other hand,  was recently 
reanalized in \cite{wong}, where a normalization  of the 
proton charge form factor  at zero momenta transfer was also treated 
as  a free parameter in the fit. The analysis of Ref. \cite{wong} leads to
a larger proton charge radius  $r_p=0.877(24)~{\rm fm}$.  
The final value for the $1$S-shift strongly depends on the 
value of the proton charge  radius:
\ba
&&\Delta E(1S)_{\rm theory} = 8~172~778(16)(32)~~{\rm kHz},
\label{theory2}
\\
&&\Delta E(1S)_{\rm theory} = 8~172~819(16)(66)~~{\rm kHz},
\label{theory3}
\ea
where the values are given for 
$r_p=0.862(12)~{\rm fm}$ and $r_p=0.877(24)~{\rm fm}$, respectively.
The first error in  all the above  equations is the theoretical uncertainty 
due to still uncalculated higher order corrections to energy levels 
beyond the $\alpha^7$ order (see a discussion below).
 The second error is due to the uncertainty
in the experimental values of the proton charge radius.
The uncertainties in Rydberg and the fine structure constants
are not relevant at the present level of precision.

Comparing these numbers  
with the  most recent measurement of the $1$S-level Lamb shift
\cite{schwob}
\be
\Delta E(1S)_{\rm exp} = 8~172~837(22)~{\rm kHz}.
\ee
we conclude that the larger values for the proton charge 
radius  seem to give an agreement between the theory and experiment.

An important part of the error in $\Delta E(1S)_{\rm theory}$  is 
due to uncalculated higher order corrections. 
For the $1S$ Lamb shift the uncertainty \cite{karsh2} 
was estimated to be about $40~{\rm kHz}$. This number 
is the linear sum of a $8~{\rm kHz}$ uncertainty  
in the self energy correction, the $16~{\rm kHz}$ uncertainty
caused by the unknown $\alpha^2(Z\alpha)^6$ terms and 
the $16~{\rm kHz}$ uncertainty due to unknown $\alpha^3 (Z\alpha)^4$
terms. The result of Ref.\cite{jent} removes the first uncertainty
and our calculation of the three loop slope of the Dirac form factor
removes the last one. The total uncertainty in the theoretical
predictions in the $1S$ Lamb shift is therefore reduced to $16~{\rm kHz}$. 

Turning the problem around, we note that the small theoretical 
uncertainty on the $1S$ Lamb  shift permits an extraction of the 
proton charge radius by comparing experimental and theoretical results.
We then arrive at the precise value of the proton radius
$r_p = 0.883 \pm 0.014~{\rm fm}$.

In conclusion, we have computed the three-loop slope of the Dirac form 
factor. Thanks to this calculation the theoretical
uncertainty in the predictions for the $1$S Lamb shift is 
reduced. Comparison of the theoretical and experimental results for the
1$S$ level shift permits an accurate determination of the proton charge 
radius. Further improvements in theoretical predictions  for the $1$S level
shift would be possible if  subleading $\alpha^2(Z\alpha)^6\log^2\alpha$
corrections are calculated.  Only then can the  theoretical 
uncertainty be brought  down to several kHz and can
 the potential of the recent measurement of the $1$S-2$S$ transition 
frequency  \cite{udem} be fully exploited. 

{\it Acknowledgments:}
We are grateful to E.~Remiddi for useful conversations and
to S.G.~Karshenboim, K.~Pachucki and A.~Yelkhovsky for comments on the 
manuscript. This research was supported in part by the United States
Department of Energy, contract DE-AC03-76SF00515, 
by BMBF under grant number BMBF-057KA92P, by
Gra\-duier\-ten\-kolleg ``Teil\-chen\-phy\-sik'' at the University of
Karlsruhe and by the DFG Forschergruppe 
``Quantenfeldtheorie, Computeralgebra und Monte-Carlo-Simulation''. 


\begin{thebibliography}{10}

\bibitem{kinoshita}
{\it Quantum Electrodynamics}, Series (World Scientific, Singapore, 1990),
edited by T. Kinoshita.


\bibitem{weitz} M.~Weitz {\it et al}, Phys. Rev. Lett. {\bf 72}, 328 (1994).

\bibitem{berkelend} D.J.~Berkeland {\it et al.}, 
Phys. Rev. Lett. {\bf 75}, 2470 (1995).

\bibitem{bourzeix} S.~Bourzeix {\it et al.}, Phys. Rev. Lett. {\bf 76},
 384 (1996).

\bibitem{beauvoir} B.~de~Beauvoir {\it et al.}  
Phys. Rev. Lett. {\bf 78}, 440 (1997).

\bibitem{udem} Th.~Udem, {\it et al.}, 
Phys. Rev. Lett. {\bf 79}, 2646 (1997).

\bibitem{schwob} C.~Schwob {\it et al.},
Phys. Rev. Lett. {\bf 82}, 4960 (1999).

\bibitem{sapirstein} J.R.~Sapirstein and D.R.~Yennie, 
  in Ref.\cite{kinoshita}, p.560.

\bibitem{mohr} P.~Mohr, in {\it Atomic, Molecular and 
Optical Physics Handbook}, edited by G.W.F.~Drake, pg. 341,
(AIP Press, Woodbory, New York, 1996).

\bibitem{preview} K.~Pachucki {\it et al.},
J. Phys.{\bf B29}, 177 (1996).

\bibitem{Karsh} S.G.~Karshenboim, Soviet Physics JETP {\bf 76}, 541 (1993).




\bibitem{ibp} 
G. 't Hooft and M. Veltman, Nucl. Phys. {\bf B44}, 189 (1972);
K.G.~Chetyrkin and F.~Tkachov, Nucl. Phys. {\bf B192}, 159 (1981).

\bibitem{Laporta} S. Laporta and E. Remiddi, Phys. Lett. {\bf B379}, 
          283 (1996).

\bibitem{brodsky} T.W.~Appelquist and S.J.~Brodsky, 
Phys. Rev. {\bf A2}, 2293 (1970).

\bibitem{barbieri} R.~Barbieri, J.A.~Magnaco and E.~Remiddi,
Nuovo Cimento Lett. {\bf 3}, 588 (1970).

\bibitem{nist} We are using the values as advocated by  National 
Institute of Standards and Technology. See http://www.nist.gov/cuu.

\bibitem{Baikov} P. Baikov and D.J. Broadhurst, hep-ph/9504398.

\bibitem{Pachucki1} K. Pachucki, Phys. Rev.{\bf A46} (1992), 648;
Ann. Phys. (N.Y.) {\bf 226}, 1 (1993).

\bibitem{Pachucki2} K.~Pachucki, Phys. Rev. Lett.{\bf 72}, 3154 (1994).

\bibitem{Eides1} M.~Eides and V.~Shelyuto, Phys. Rev.{\bf A52}, 954 (1995).

\bibitem{jent} U.D.~Jentschura, P.J.~Mohr and G.~Soff,
Phys. Rev. Lett. {\bf 82}, 53 (1999).

 
\bibitem{Salpeter} E.E.~Salpeter, Phys. Rev. {\bf 87}, 328 (1952).

\bibitem{Pachucki3} K.~Pachucki and H.~Grotch, Phys. Rev. {\bf A51}, 1854
 (1995).
 
\bibitem{Pachucki4}  K.~Pachucki, Phys. Rev. {\bf A52}, 1079 (1995).

\bibitem{friar} J.L.~Friar {\it et al.}, Phys. Rev. {\bf A59}, 4061 (1999).

\bibitem{hand} L.N. Hand {\it et al.}, Rev. Mod. Phys. {\bf 35}, 335 (1963). 

\bibitem{simon} G.G.~Simon {\it et al.}, 381 (1980).  

\bibitem{wong} Ch.W.~Wong, Int. J. Mod. Phys. {\bf 3}, 821 (1994).

\bibitem{karsh2} S.G.~Karshenboim, Can. J. Phys. {\bf 77}, 241 (1999).

\end{thebibliography}

\end{document}